\documentclass[twocolumn,reprint,amsmath,amssymb, aps,prl]{revtex4-2}
\usepackage[T1]{fontenc}
\usepackage[latin9]{inputenc}
\makeatletter
\usepackage{float}
\usepackage{physics}
\usepackage{amsfonts}
\usepackage{latexsym}
\usepackage{epsfig}
\usepackage{graphicx}
\usepackage{tikz}
\definecolor{greatblue}{RGB}{40,120,181}
\definecolor{greatred}{RGB}{200,36,35}
\usepackage[colorlinks,linkcolor=greatblue,anchorcolor=blue,citecolor=greatred]{hyperref}

\usepackage{dcolumn}% Align table columns on decimal point
\usepackage{bm}% bold math

\makeatother

\usepackage[english]{babel}

\begin{document}
\preprint{preprintnumbers}{CTP-SCU/2025012}
\title{Derive Einstein equation  from CFT entanglement entropy}

\author{Xin Jiang}
\email{domoki@stu.scu.edu.cn}
\affiliation{College of Physics, Sichuan University, Chengdu, 610065, China}

\author{Haitang Yang}
\email{hyanga@scu.edu.cn}
\affiliation{College of Physics, Sichuan University, Chengdu, 610065, China}

%\date{\today}

\begin{abstract}
We explicitly show how to derive the $(D+1)$-dimensional Einstein equation from the   entanglement entropy 
between codimension-one {\it disjoint}  regions  in $D$-dimensional conformal field theory. 

%For all dimension $D$, using the {\it finite}  entanglement entropy  between disjoint codimension-one regions, we  demonstrate
%how $(D+1)$-dimensional Einstein equation emerges from   CFT$_D$ entanglement entropies.
\end{abstract}

\maketitle

%\onecolumngrid

\section{Introduction}

Over the past few decades, our understanding of gravity's nature has undergone profound paradigm shifts. 
This transformation is initiated by the Gauge/Gravity duality, 
also known as the AdS/CFT correspondence \citep{Maldacena:1997re,Gubser:1998bc,Witten:1998qj}, which
suggests an equivalence between  a $(D+1)$-dimensional gravitational theory in AdS and a  
$D$-dimensional QFT  living on its boundary. 

Motived by this duality and the Bekenstein-Hawking entropy \citep{Bekenstein:1973ur,Hawking:1975vcx},
Ryu and Takayanagi proposed a landmark identification between the entanglement entropy of CFT$_D$
and the area of the corresponding minimal surface in AdS$_{D+1}$ \citep{Ryu:2006bv,Ryu:2006ef}.  
For a  CFT$_D$ partitioned into {\it adjacent and complementary} regions $A$ and $A^c$, 
the entanglement entropy is defined by the von Neumann entropy 
$S_{\rm adj}(A:A^c)=-\mathrm{Tr}\,\rho_{A}\log\rho_{A}$, 
where $\rho_A = \mathrm{Tr}_{A^c}\rho$ is the reduced density matrix for region $A$.  
The subscript ``adj'' indicates adjacent entangling regions.  
The AdS$_{D+1}$ counterpart of $S_{\rm adj}(A:A^c)$ is a codimension-two hypersurface homologous to $A$.
 
The RT formula, along with developments such as ``ER=EPR''  conjecture \citep{Maldacena:2013xja},  
has since motivated a profound interpretation: the nature of 
spacetime is quantum entanglement \cite{VanRaamsdonk:2010pw}. 
However,  a  derivation of the  Einstein equation from quantum entanglement 
has remained unresolved over a long time.  
This problem stems from two critical challenges:
\begin{enumerate}
\item The entanglement entropy $S_{\rm adj}(A:A^c)$ between adjacent regions is divergent,
arising from the  intense short-range correlations in contiguous fields.
\item Until recently, no general method existed to compute 
entanglement entropies in CFT$_D$ for $D\ge 3$. 
\end{enumerate}    
The first issue makes the original RT formula an asymptotic relation,
rendering  it  extremely challenging, if not completely impossible, 
to extract exact relations between gravity and CFT.
To establish an exact RT formula, there is a need for 
finite {\it elementary} entanglement entropies, 
which are intended to measure the entanglement between {\it disjoint} subregions,
as opposed to adjacent ones.
Here, the term ``elementary'' is used in contrast
to compound quantities such as mutual information, relative entropy, and the like --- which are
constructed from combinations of elementary entanglement measures.

There are two additional intrinsic reasons to study the entanglement between disjoint subregions in CFT.
\begin{enumerate}
\item  For UV-complete theories,  
UV divergences should not be intrinsic ingredients 
but be  UV limits of  regular quantities. 
Since an adjacent configuration is obviously an adjacent limit of a disjoint configuration, 
and the entanglement entropy between disjoint regions should be finite,
investigating disjoint configurations becomes a necessity.
\item Another critical reason  comes from 
the Reeh-Schlieder theorem which demonstrates that any two subregions in a QFT are 
entangled \cite{Reeh:1961ujh,Witten:2018zxz}.
\end{enumerate}
In prior collaborations \citep{Jiang:2024ijx,Jiang:2025tqu,Jiang:2025dir},  
we showed that  the entanglement entropies $S_\mathrm{disj}(A:B)$ 
(subscript ``disj'' denoting the disjoint configuration) 
between disjoint intervals $A$ and $B$ in CFT$_2$ are 
indeed finite, in static, covariant and 
thermal settings. 
The divergent adjacent $S_{\rm adj}(A:B)$ arises as the simple adjacent limit of 
the disjoint $S_\mathrm{disj}(A:B)$. 
The finiteness of disjoint $S_\mathrm{disj}(A:B)$ enables
us  to explicitly derive the three dimensional Einstein equation
from CFT$_2$  entanglement entropy  in \citep{Jiang:2025isp}. 

However, the three dimensional gravity is topological and has no 
local dynamical degrees of freedom, 
making higher dimensional studies critical for uncovering the quantum origins of gravity. 
Yet this remained intractable due to the second challenge until our 
recent work \citep{Jiang:2025jnk}, in which we introduced a systematic 
field theory method to compute the entanglement entropy of CFT$_D$ 
in all dimensions.
With this advancement, we are going to complete 
the derivation of the $(D+1)$-dimensional
Einstein equation from the  entanglement entropy of CFT$_D$ for $D\ge 2$ 
in this work.

\section{Entanglement entropy for disjoint subsystems in CFT$_{D}$}

We begin by briefly outlining the calculation of entanglement entropy 
for $\text{CFT}_D$ ($D \geq 2$) given in  \citep{Jiang:2025jnk}.
As we explained in the introduction, in order to derive exact relations such as Einstein equation,
we need to find the finite entanglement entropy $S_\mathrm{disj}(A:B)$, which measures 
the entanglement between disjoint regions $A$ and $B$ in CFT$_D$. 
This is also a direct consequence of the Reeh-Schlieder theorem \cite{Reeh:1961ujh,Witten:2018zxz},
which shows that any two subregions in a QFT are inherently entangled.
The usual divergent 
$S_\mathrm{adj}(A:B)$ which quantifies  the entanglement between adjacent regions
is a simple adjacent limit of the disjoint  $S_\mathrm{disj}(A:B)$.

\begin{figure}[h]
\centering
\includegraphics[scale=0.4]{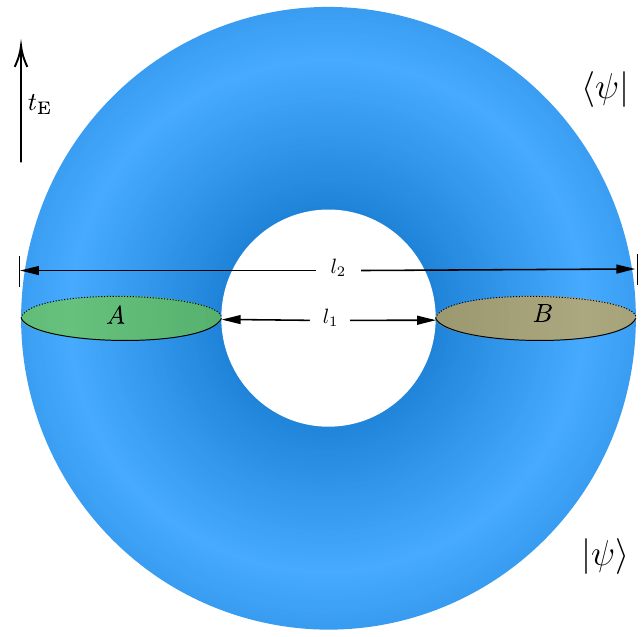}
\caption{The pure state density matrix $\rho=|\psi\rangle\langle\psi|$ of the solid torus CFT$_D$. 
Entangling regions $A$ and $B$ (colored)  are disjoint, ensuring a finite
entanglement entropy between $A$ and $B$.
\label{fig:solid-torus}}
\end{figure}

Consider a $\text{CFT}_D$ in Euclidean spacetime with flat metric 
$\mathrm{d}s_{0}^{2} = \mathrm{d}t_{\text{E}}^{2} + \mathrm{d}y^{2} + \sum_{I=1}^{D-2}\mathrm{d}x_{I}^{2}$. 
To ensure a finite entanglement entropy $S_\text{disj}(A:B)$, two conditions must hold:
\begin{enumerate}
    \item $A$ and $B$ are disjoint regions;
    \item $A$ and $B$ are complementary, forming a pure state.
\end{enumerate}
The solid torus shown in Fig. \ref{fig:solid-torus} precisely satisfies these two conditions.
Note the time direction is upward. This geometry represents quantum fields living in  a bounded region 
\begin{equation}
\mathcal{B} := \left\{ \left(\sqrt{t_{\text{E}}^{2} + y^{2}} - \frac{l_{2} + l_{1}}{2}\right)^{2} + \sum_{I=1}^{D-2} x_{I}^{2} < \left(\frac{l_{2} - l_{1}}{2}\right)^{2} \right\},
\end{equation}
with $l_{2} > l_{1} > 0$. At $t_{\text{E}} = 0$, $\mathcal{B}$ consists 
of  two disjoint $(D-1)$-balls $A$ and $B$ of equal radius.

The entanglement entropy $S_\mathrm{disj}(A:B)$ can be computed via the replica 
trick \citep{Callan:1994py,Holzhey:1994we}:
\begin{equation}
S_\mathrm{disj}(A:B) = \lim_{n \rightarrow 1} \frac{1}{1 - n} \log \mathrm{Tr}_A \rho_A^n,
\end{equation}
where $\rho_A = \mathrm{Tr}_B \rho$ and 
$$\mathrm{Tr}_A \rho_A^n = \frac{Z_n}{Z_1^n}.$$
Here, $Z_1$ 
is the partition function of the solid torus $\mathcal{B}$. 
$Z_n$ is the partition function on the $n$-sheeted 
cover $\mathcal{B}_n$, obtained by cyclically gluing $n$ copies of $\mathcal{B}$  
along $A$. 
Clearly, $\mathcal{B}_n$ remains a solid torus, but with a period of $2n\pi$.
So, the calculation of the disjoint entanglement entropy is simplified to
computing the partition function of the solid torus in Fig. \ref{fig:solid-torus},
which leads to \citep{Jiang:2025jnk}, 
\begin{equation}
S_\mathrm{disj}(A:B) = \gamma \left(\frac{l_{2}}{l_{1}} - 1\right)^{D-1} \,_{2}F_{1}\left(D-1, 
\frac{D}{2}; D; 1 - \frac{l_{2}}{l_{1}}\right),
\label{eq:S-bulk}
\end{equation}
where $\gamma$ is a model-dependent constant and $\,_{2}F_{1}$ denotes the hypergeometric function. 

% 
%The entanglement entropy $S_\mathrm{disj}(A:B)$ for
%$A$ and $B$ with different radii and time  can be 
%obtained with conformal transformations. 

\begin{figure}[h]
\begin{centering}
\includegraphics[scale=0.6]{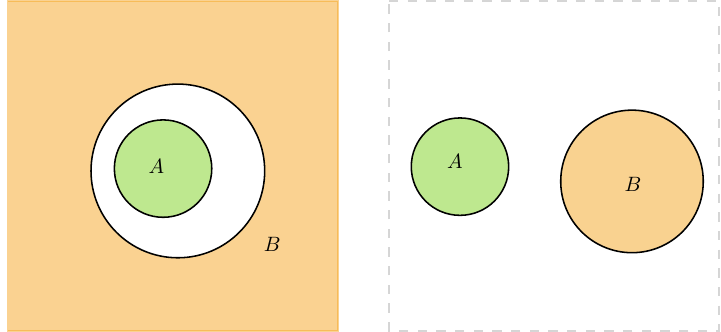}
\par\end{centering}
\caption{Two configurations are conformally equivalent to the solid torus. Left panel (the cavity configuration):
A ($D-1$)-ball living in a spheric cavity is entangled with the region
outside the cavity. Right panel (the juxtaposed configuration): Two disjoint
($D-1$)-balls with different radii are entangled.
\label{fig:two-config}}
\end{figure}

In the solid torus, $A$ and $B$  are symmetric and share the same radius.
For $A$ and $B$ having different radii,  
there are two configurations, namely juxtaposed configuration and cavity  configuration,
as illustrated in Fig. \ref{fig:two-config}. 
Both configurations are conformally equivalent to 
the solid torus and their entanglement entropy $S_\mathrm{disj}(A:B)$
can be obtained with conformal transformations from eq. (\ref{eq:S-bulk}).

\begin{figure}[H]
\centering
\includegraphics[scale=0.8]{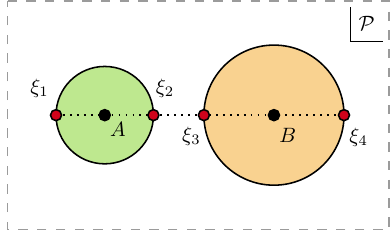}
\caption{Two disjoint $(D-1)$-balls $A$ and $B$ on a spacelike hyperplane $\mathcal{P}$, determined by four collinear endpoints (red dots). The cross ratio $\eta$ (Eq. \ref{eq:cross-ratio}) parametrizes their relative positions, governing the entanglement entropy via conformal invariance.\label{fig:Entangling-regions}}
\end{figure}

Obviously, two balls has $2(D+1)$ degrees of freedom, fixing
the centers and radii.
For our purpose, 
it suffices to consider that $A$ and $B$ are living
on a Euclidean hyperplane $\mathcal{P}$, and are uniquely fixed by
four collinear points $\xi_{I}=(t_{I},\vec{x}_{I}),I=1,\dots,4$,
as illustrated in Figure \ref{fig:Entangling-regions}. 
Note that although each of $\xi_I$'s has $D$ components, 
there are really $2(D+1)$ degrees of freedom: $D$ for $\xi_1$, $1$ for 
each of $\xi_2$, $\xi_3$,
$\xi_4$, and $(D-1)$ for rotational symmetry.
The real time
follows from a Wick rotation $t_{\text{E}}\rightarrow\mathrm{i}t$,
where the hyperplane $\mathcal{P}$ should be always spacelike. 
The  disjoint entanglement entropy  $S_\mathrm{disj}(A:B)$ for generic
configurations  can be easily obtained by writing $l_2/l_1$ in eq. (\ref{eq:S-bulk}) 
in terms of the conformal invariant cross-ratio $\eta$, 
\begin{equation}
\frac{l_2}{l_1}= \frac{1+\eta^{1/2}}{1-\eta^{1/2}}, \quad
\eta \equiv \frac{\vert\xi_{12}\vert\vert\xi_{34}\vert}{\vert\xi_{13}\vert\vert\xi_{24}\vert},
\label{eq:cross-ratio}
\end{equation}
with $\vert\xi_{IJ}\vert \equiv |\xi_I -\xi_J| \equiv \sqrt{(\vec{x}_I - \vec{x}_J)^2 - (t_I - t_J)^2}$ 
denoting the Lorentzian distance between points $\xi_I$ and $\xi_J$. 
So, for generic configurations, we get \citep{Jiang:2025jnk},
%\begin{widetext} 
%\begin{equation}
%S_\mathrm{disj}(A:B) = \gamma \left(\frac{2}{\eta^{-\frac{1}{2}} - 1}\right)^{D-1} 
%\,_{2}F_{1}\left(D-1, \frac{D}{2}; D; \frac{-2}{\eta^{-\frac{1}{2}} - 1}\right).
%\label{eq:S-CFT}
%\end{equation}
%\end{widetext}
%where the cross ratio $\eta$ is
%\begin{equation}
%\eta = \frac{\vert\xi_{12}\vert\vert\xi_{34}\vert}{\vert\xi_{13}\vert\vert\xi_{24}\vert},
%\label{eq:cross-ratio}
%\end{equation}
%This expression highlights that $S_\mathrm{disj}(A:B)$ is indeed finite.
%\begin{widetext} 
\begin{equation}
S_\mathrm{disj}(A:B) = \gamma \left(\frac{2\eta^{\frac{1}{2}}}{1-\eta^{\frac{1}{2}}}\right)^{D-1} 
\,_{2}F_{1}\left(D-1, \frac{D}{2}; D; \frac{-2\eta^{\frac{1}{2}}}{1-\eta^{\frac{1}{2}}}\right).
\label{eq:S-CFT}
\end{equation}
%\end{widetext}
%This is the general result for all disjoint configurations,
%although we used the juxtaposed configuration as an illustration. 

\begin{figure}[h]
\begin{centering}
\includegraphics[scale=0.6]{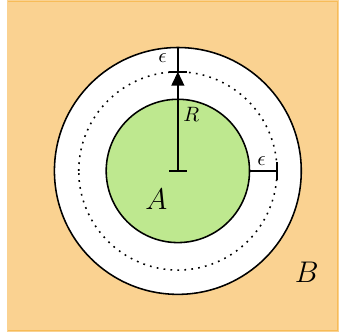}
\par\end{centering}
\caption{As $\epsilon\to 0$, the cavity configuration reduces to the usual
adjacent configuration. 
\label{fig:UV-case}}
\end{figure}

It is clear that the usual adjacent configuration is a simple limit $\epsilon\to 0$ of the 
cavity configuration as shown in Fig. \ref{fig:UV-case}. In \citep{Jiang:2025jnk},
explicit expressions for $D=2$ to $6$ are presented by expanding eq. (\ref{eq:S-CFT})
and they perfectly match the holographic results given in  \citep{Ryu:2006bv,Ryu:2006ef}.

\section{Einstein equation from entanglement entropies in CFT$_{D}$}

The parameterization $\xi_I$ in eq. (\ref{eq:cross-ratio}) and Fig. \ref{fig:Entangling-regions}
is chosen for  convenience.
We certainly can use whatever equivalent parameterization to express  $S_\mathrm{disj}(A:B)$.
For example, we could alternatively adopt the radii and centers of $A$ and $B$ as parameters.
In particular, for the symmetric case where $A$ and $B$
share the same radius (corresponding to the solid torus depicted in Fig. \ref{fig:solid-torus}), 
we now select the following linear combination as an alternative parameterization,
\begin{eqnarray}
X &=& \frac{\xi_4 +\xi_1}{2},\quad {X'} = \frac{\xi_3 +\xi_2}{2},\nonumber\\
Z &=& \frac{|\xi_4 -\xi_1|}{2},\quad {Z'} = \frac{|\xi_3 -\xi_2|}{2}.
\label{eq:sysRe}
\end{eqnarray}
For a generic configuration that $A$ and $B$ having different radii, 
such as that in  Fig.   \ref{fig:Entangling-regions}, 
the  reparameterizations are derived by applying the  
corresponding conformal transformation to Eq. (\ref{eq:sysRe}), which yields,
\begin{eqnarray}
X & = & \xi_{1}+\frac{1}{1+\alpha}\xi_{41},\quad {X^{\prime}}  =  \xi_{2}+\frac{1}{1+\beta}\xi_{32},\\
Z & = & \frac{\sqrt{\alpha}}{1+\alpha}\vert\xi_{14}\vert,\quad\quad Z^{\prime}  =  \frac{\sqrt{\beta}}{\beta+1}\vert\xi_{23}\vert,
\end{eqnarray}
with $\alpha=\frac{\vert\xi_{34}\vert\vert\xi_{24}\vert}{\vert\xi_{13}\vert\vert\xi_{12}\vert}$
and $\beta=\frac{\vert\xi_{13}\vert\vert\xi_{34}\vert}{\vert\xi_{12}\vert\vert\xi_{24}\vert}$.
In terms of the new parameters,  the conformal invariant cross-ratio  is
\begin{equation}
\eta = \frac{\vert\xi_{12}\vert\vert\xi_{34}\vert}{\vert\xi_{13}\vert\vert\xi_{24}\vert}=
\frac{(Z-Z')^{2}+\vert X-X'\vert^{2}}{(Z+Z')^{2}+\vert X-X'\vert^{2}}.\label{eq:cross-ratio-x}
\end{equation}
We then introduce a nontrivial grouping: $Y^\mu \equiv (X,Z)$ and ${Y'}^\mu\equiv (X',Z')$.
Note that $Y^\mu, {Y'}^\mu$ could be viewed as points in a ($D+1$)-dimensional spacetime, 
since the dimensionality of $X$ or $X'$ is $D$. 
A $(D+1)$-dimensional metric thus can be defined
\begin{eqnarray}
\chi(Y,Y')&:=&\frac{D-1}{2} \left( G^{(D+1)} S_{\mathrm{disj}}(A:B)\right)^{\frac{2}{D-1}},\nonumber\\
g_{\mu\nu}&=& -[\chi_{\mu\nu'}]:=-\lim_{Y'\to Y}\partial_{\nu'}\partial_{\mu}\chi,\label{eq:chi}
\end{eqnarray}
where $ G^{(D+1)}$ is the $(D+1)$-dimensional Newton constant.
The bracket $[\cdots]$ denotes the coincidence limit $Y'\to Y$.
Indices in brackets hereinafter are derivatives with respect to $Y^{\mu}$ or
$Y'^{\mu}$, $\mu=1,\dots,D+1$. Introduce 

\begin{equation*}
R_{\mu\nu}:=   -[\chi^{\rho\sigma'}\, \nabla_\nu\nabla_\sigma\nabla_\mu\partial_\rho \chi],
\end{equation*}
\begin{equation}
R:= -[\chi^{\mu\nu'}] R_{\mu\nu} =  
[ \chi^{\mu\nu'}\chi^{\rho\sigma'}\, \nabla_\nu\nabla_\sigma\nabla_\mu\partial_\rho \chi],
\label{eq:definitions}
\end{equation}
%
%\begin{equation}
%R_{\mu\nu}:=g^{\rho\sigma}[\nabla_{\nu}\nabla_{\sigma}\nabla_{\mu}\partial_{\rho}\chi],\quad R:=g^{\mu\nu}R_{\mu\nu},\label{eq:definitions}
%\end{equation}
%where the covariant derivatives $\nabla_{(\centerdot)}$ are defined by $g_{\mu\nu}$. 
where the covariant derivatives  $\nabla_{(\centerdot)}$  are defined by 
$g_{\mu\nu}\equiv -[\chi_{\mu\nu'}]$ in eq. (\ref{eq:chi}).
Substituting the explicit form of $\chi$ in equations (\ref{eq:S-CFT}), (\ref{eq:cross-ratio-x}) 
and (\ref{eq:chi})  into these definitions, 
we obtain after straightforward but lengthy calculations:
\begin{equation}
R_{\mu\nu}-\frac{1}{2}g_{\mu\nu}R +\Lambda g_{\mu\nu}=0,\label{eq:EE-EOM}
\end{equation}
with the cosmological constant
\begin{equation}
\Lambda = \frac{D-1}{2(D+1)} R = \frac{D-1}{2(D+1)} 
\Big[\chi^{\mu\nu'}\chi^{\rho\sigma'}\nabla_\nu\nabla_\sigma\nabla_\mu\partial_\rho \chi\Big].
\label{eq:Lambda}
\end{equation}
This is precisely the ($D+1$)-dimensional Einstein equation! 

Several fundamental insights emerge from our derivation:
\begin{itemize}
    \item The ($D+1$)-dimensional Einstein equation emerges solely from the entanglement structure of a $D$-dimensional CFT.
    \item From eq. (\ref{eq:chi}), the metric is a derived but not a fundamental object, why bother quantize it?
    \item Unlike conventional gravity theories, $\Lambda$ is not an input parameter but is derived from the CFT data. 
\end{itemize}

\section{Intuitive understanding of the derivation}

It might appear mysterious why and how the derivation works. 
Thus, while we have derived the Einstein equation purely from CFT data, 
it is instructive to provide an intuitive holographic interpretation in this section.

The RT formula tells us that the gravitational dual of  CFT 
entanglement entropy is a codimension-two minimal surface in the bulk. 
Since we here find finite entanglement entropy $S_\text{disj}(A:B)$,
the equality of $S_\text{disj}(A:B)$ and RT surface should be exact, 
as proposed in \citep{Takayanagi:2017knl}.

\begin{figure}[h]
\includegraphics[width=0.35\textwidth]{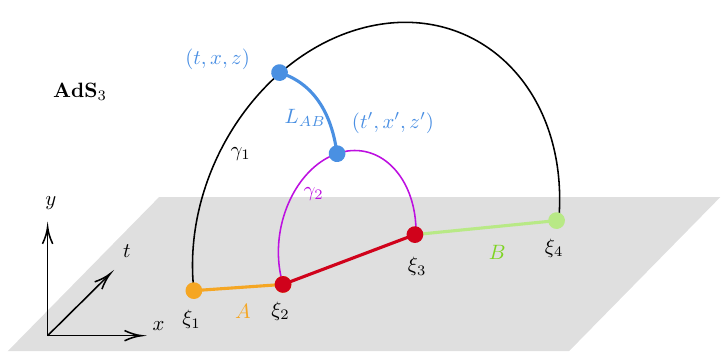}
\caption{In AdS$_3$/CFT$_2$, the RT surfaces are geodesics. The geodesic dual to 
$S_\text{disj}(A:B)$ is $L_{AB}$.
\label{fig:geodesic}}
\end{figure}

For $D=2$ (AdS$_3$/CFT$_2$), the minimal surfaces reduce to geodesics,
as depicted in Fig. \ref{fig:geodesic}. 
Remarkably, the metric can be directly extracted  from the geodesic length $L(Y,Y')$
between points $Y$ and $Y'$  \citep{Synge:1960ueh},
\begin{equation}
g_{\mu\nu}= -\lim_{Y'\to Y} \partial_{\nu'} \partial_{\mu} 
\left(\frac{1}{2} L^2(Y,Y')\right).
\label{eq:GtoM1}
\end{equation}
By using this equation, in \citep{Jiang:2025isp}, we successfully
derived $3D$ Einstein equation from CFT$_2$ entanglement entropy.

But how  does this generalize to  $D\ge 3$ in which minimal surfaces are no longer geodesics?
Notably, there is mathematically no general method to extract a metric from  minimal surfaces
in higher dimensions.
We can understand the mechanism as follows. 

\begin{figure}[h]
\includegraphics[width=0.35\textwidth]{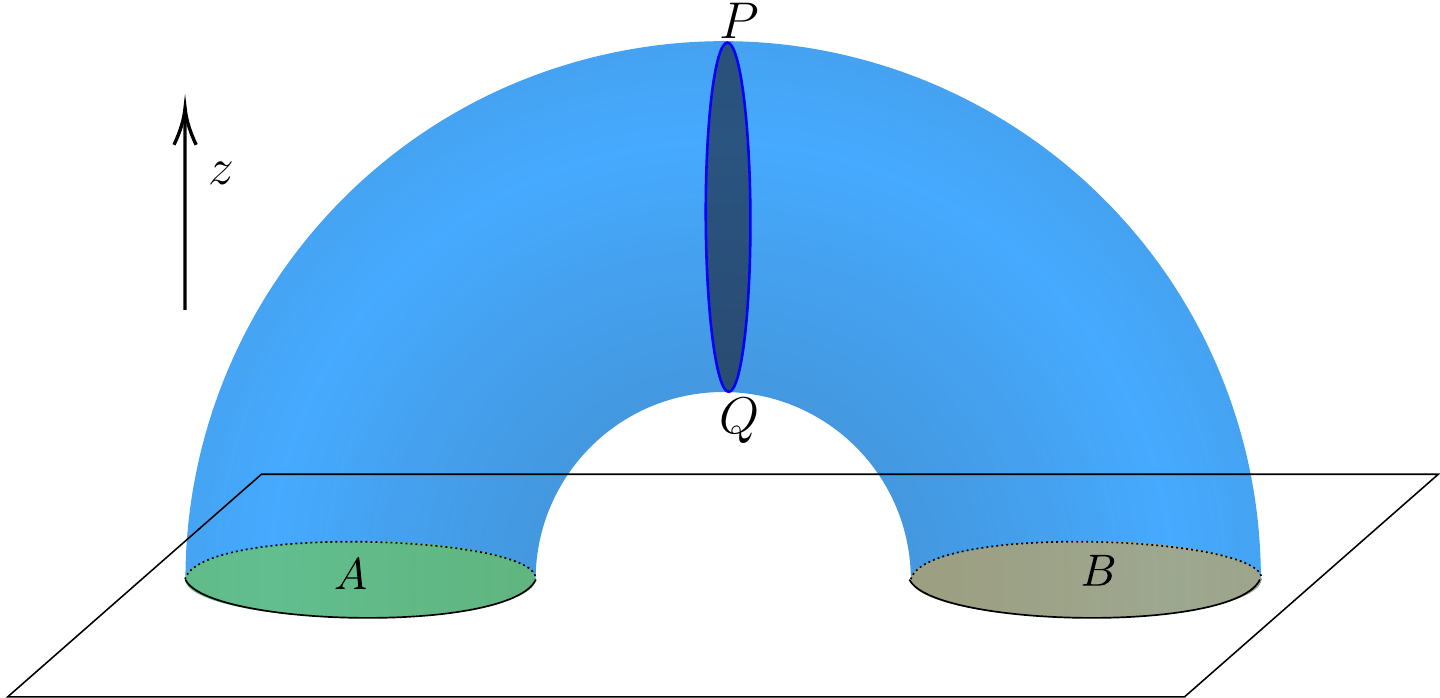}
\caption{Ball $A$ provides $D+1$ parameters by its center and radius. So does ball $B$.
These $2(D+1)$ parameters yield exactly the degrees of freedom needed to define two antipodal 
points $P$ and $Q$ in AdS$_{D+1}$.
Points $P$ and $Q$ fix a geodesic, which  subsequently fixes a minimal disc 
up to rotational gauge redundancy, which in turn fixes a minimal
ball  up to rotational gauge redundancy, and so on.
\label{fig:illustration}}
\end{figure}

The crucial insight is that the entanglement entropy $S_{\mathrm{disj}}(A:B)$ of CFT$_D$ 
is fully  determined  by $2(D+1)$ parameters: the centers and  radii  of balls $A$ and $B$. 
These $2(D+1)$ parameters provide the exact degrees of freedom to fix 
two antipodal points $P$ and $Q$ in a $(D+1)$-dimensional spacetime,
as shown in Fig. \ref{fig:illustration}.

Points $P$ and $Q$ fix a geodesic, which in turn  
fixes a minimal disc up to a gauge freedom, which subsequently fixes a minimal
ball up to a gauge freedom, and so on,
The gauge freedoms justify choosing collinear $\xi_i$'s to derive eq. (\ref{eq:S-CFT}).

In short, the  RT surface in AdS$_{D+1}$  actually is fixed by 
a geodesic between two points, as indicated in eq. (\ref{eq:S-bulk}),
where we can see that $S_{\mathrm{disj}}(A:B)$ is a function of 
the dimensionality $D$ and geodesic length $L(P,Q)= \log(l_2/l_1)$.
So, what we need to do is to isolate the geodesic from the minimal surface.
From eq. (\ref{eq:chi}),  taking  the double derivatives and the limit $Y'\to Y$  isolates
the quadratic term $\Delta Y^\mu \Delta Y^\nu$ in the entanglement entropy expansion.

The exponential in $ \left( G^{(D+1)} S_{\mathrm{disj}}(A:B)\right)^{\frac{2}{D-1}}$ can be easily understood by dimensional analysis since the holographic dual of 
entanglement entropy is a codimension-two surface.

\section{Summary and Discussions}

In this paper, we derived the $(D+1)$-dimensional vacuum Einstein equation from 
the disjoint entanglement entropy $S_{\mathrm{disj}}(A:B)$ of $D$-dimensional CFT. 
The finiteness of $S_{\mathrm{disj}}(A:B)$ is essential  to the derivation. 

In a previous collaboration \citep{Jiang:2025isp}, we derived $3D$ Einstein equation from
the disjoint entanglement entropy of CFT$_2$. Since $3D$ gravity is locally trivial, 
it is worthwhile to restate some important remarks. 
Referring to eq. (\ref{eq:chi}), the metric is given by the entanglement entropy of CFT$_D$.
So,
\begin{itemize}
\item The metric is not a fundamental object but is derived from a quantum quantity,
suggesting that quantizing the metric may not be the right way to addressing  ``quantum gravity''.
\item The non-local/local duality between boundary and bulk is explicit, 
as  entanglement entropy is a non-local quantity in CFT.
\item When $S_\mathrm{disj}(A:B)$ is non-analytic (e.g., during phase transitions), 
bulk metric singularities emerge. Classifying these singularities may offer insights 
into black hole physics and quantum critical phenomena.
\end{itemize} 
For the future research, we think the most important direction is to derive the matter sourced 
Einstein equation. This should be a highly nontrivial task since the conformal symmetry no longer exists.

%Moreover, the RG equation is nothing but a geometric identity in CFT$_2$. This also holds for CFT$_D$, $D\ge 3$.

\vspace*{3.0ex}
\begin{acknowledgments}
\paragraph*{Acknowledgments.} 
This work is supported in part by NSFC (Grant No. 12275184).
\end{acknowledgments}

\bibliographystyle{unsrturl}
\bibliography{ref202510}

\begin{thebibliography}{10}

\bibitem{Maldacena:1997re}
Juan~Martin Maldacena.
\newblock {The Large N limit of superconformal field theories and
  supergravity}.
\newblock {\em Adv. Theor. Math. Phys.}, 2:231--252, 1998.
\newblock \href {http://arxiv.org/abs/hep-th/9711200}
  {\path{arXiv:hep-th/9711200}}, \href
  {http://dx.doi.org/10.1023/A:1026654312961}
  {\path{doi:10.1023/A:1026654312961}}.

\bibitem{Gubser:1998bc}
S.~S. Gubser, Igor~R. Klebanov, and Alexander~M. Polyakov.
\newblock {Gauge theory correlators from noncritical string theory}.
\newblock {\em Phys. Lett. B}, 428:105--114, 1998.
\newblock \href {http://arxiv.org/abs/hep-th/9802109}
  {\path{arXiv:hep-th/9802109}}, \href
  {http://dx.doi.org/10.1016/S0370-2693(98)00377-3}
  {\path{doi:10.1016/S0370-2693(98)00377-3}}.

\bibitem{Witten:1998qj}
Edward Witten.
\newblock {Anti-de Sitter space and holography}.
\newblock {\em Adv. Theor. Math. Phys.}, 2:253--291, 1998.
\newblock \href {http://arxiv.org/abs/hep-th/9802150}
  {\path{arXiv:hep-th/9802150}}, \href
  {http://dx.doi.org/10.4310/ATMP.1998.v2.n2.a2}
  {\path{doi:10.4310/ATMP.1998.v2.n2.a2}}.

\bibitem{Bekenstein:1973ur}
Jacob~D. Bekenstein.
\newblock {Black holes and entropy}.
\newblock {\em Phys. Rev. D}, 7:2333--2346, 1973.
\newblock \href {http://dx.doi.org/10.1103/PhysRevD.7.2333}
  {\path{doi:10.1103/PhysRevD.7.2333}}.

\bibitem{Hawking:1975vcx}
S.~W. Hawking.
\newblock {Particle Creation by Black Holes}.
\newblock {\em Commun. Math. Phys.}, 43:199--220, 1975.
\newblock [Erratum: Commun.Math.Phys. 46, 206 (1976)].
\newblock \href {http://dx.doi.org/10.1007/BF02345020}
  {\path{doi:10.1007/BF02345020}}.

\bibitem{Ryu:2006bv}
Shinsei Ryu and Tadashi Takayanagi.
\newblock {Holographic derivation of entanglement entropy from AdS/CFT}.
\newblock {\em Phys. Rev. Lett.}, 96:181602, 2006.
\newblock \href {http://arxiv.org/abs/hep-th/0603001}
  {\path{arXiv:hep-th/0603001}}, \href
  {http://dx.doi.org/10.1103/PhysRevLett.96.181602}
  {\path{doi:10.1103/PhysRevLett.96.181602}}.

\bibitem{Ryu:2006ef}
Shinsei Ryu and Tadashi Takayanagi.
\newblock Aspects of holographic entanglement entropy.
\newblock {\em Journal of High Energy Physics}, 2006(08):045--045, aug 2006.
\newblock \href {http://dx.doi.org/10.1088/1126-6708/2006/08/045}
  {\path{doi:10.1088/1126-6708/2006/08/045}}.

\bibitem{Maldacena:2013xja}
Juan Maldacena and Leonard Susskind.
\newblock {Cool horizons for entangled black holes}.
\newblock {\em Fortsch. Phys.}, 61:781--811, 2013.
\newblock \href {http://arxiv.org/abs/1306.0533} {\path{arXiv:1306.0533}},
  \href {http://dx.doi.org/10.1002/prop.201300020}
  {\path{doi:10.1002/prop.201300020}}.

\bibitem{VanRaamsdonk:2010pw}
Mark Van~Raamsdonk.
\newblock {Building up spacetime with quantum entanglement}.
\newblock {\em Gen. Rel. Grav.}, 42:2323--2329, 2010.
\newblock \href {http://arxiv.org/abs/1005.3035} {\path{arXiv:1005.3035}},
  \href {http://dx.doi.org/10.1142/S0218271810018529}
  {\path{doi:10.1142/S0218271810018529}}.

\bibitem{Reeh:1961ujh}
H.~Reeh and S.~Schlieder.
\newblock {Bemerkungen zur unit{\"a}r{\"a}quivalenz von lorentzinvarianten
  feldern}.
\newblock {\em Nuovo Cim.}, 22(5):1051--1068, 1961.
\newblock \href {http://dx.doi.org/10.1007/BF02787889}
  {\path{doi:10.1007/BF02787889}}.

\bibitem{Witten:2018zxz}
Edward Witten.
\newblock {APS Medal for Exceptional Achievement in Research: Invited article
  on entanglement properties of quantum field theory}.
\newblock {\em Rev. Mod. Phys.}, 90(4):045003, 2018.
\newblock \href {http://arxiv.org/abs/1803.04993} {\path{arXiv:1803.04993}},
  \href {http://dx.doi.org/10.1103/RevModPhys.90.045003}
  {\path{doi:10.1103/RevModPhys.90.045003}}.

\bibitem{Jiang:2024ijx}
Xin Jiang, Peng Wang, Houwen Wu, and Haitang Yang.
\newblock {Alternative to purification in conformal field theory}.
\newblock {\em Phys. Rev. D}, 111(2):L021902, 2025.
\newblock \href {http://arxiv.org/abs/2406.09033} {\path{arXiv:2406.09033}},
  \href {http://dx.doi.org/10.1103/PhysRevD.111.L021902}
  {\path{doi:10.1103/PhysRevD.111.L021902}}.

\bibitem{Jiang:2025tqu}
Xin Jiang, Peng Wang, Houwen Wu, and Haitang Yang.
\newblock {Mixed state entanglement entropy in CFT}.
\newblock {\em JHEP}, 09:133, 2025.
\newblock \href {http://arxiv.org/abs/2501.08198} {\path{arXiv:2501.08198}},
  \href {http://dx.doi.org/10.1007/JHEP09(2025)133}
  {\path{doi:10.1007/JHEP09(2025)133}}.

\bibitem{Jiang:2025dir}
Xin Jiang, Haitang Yang, and Zilin Zhao.
\newblock {Entanglement entropy of mixed state in thermal CFT2}.
\newblock {\em Phys. Rev. D}, 112(4):046025, 2025.
\newblock \href {http://arxiv.org/abs/2501.11302} {\path{arXiv:2501.11302}},
  \href {http://dx.doi.org/10.1103/bpzx-kdgq} {\path{doi:10.1103/bpzx-kdgq}}.

\bibitem{Jiang:2025isp}
Xin Jiang, Peng Wang, Houwen Wu, and Haitang Yang.
\newblock {How Einstein{\textquoteright}s equations emerge from CFT$_2$}.
\newblock {\em Phys. Rev. D}, 112(8):L081906, 2025.
\newblock \href {http://arxiv.org/abs/2410.19711} {\path{arXiv:2410.19711}},
  \href {http://dx.doi.org/10.1103/zg5x-34mn} {\path{doi:10.1103/zg5x-34mn}}.

\bibitem{Jiang:2025jnk}
Xin Jiang and Haitang Yang.
\newblock {Entanglement Entropy of Conformal Field Theory in All Dimensions}.
\newblock 6 2025.
\newblock \href {http://arxiv.org/abs/2506.02786} {\path{arXiv:2506.02786}}.

\bibitem{Callan:1994py}
Curtis~G. Callan, Jr. and Frank Wilczek.
\newblock {On geometric entropy}.
\newblock {\em Phys. Lett. B}, 333:55--61, 1994.
\newblock \href {http://arxiv.org/abs/hep-th/9401072}
  {\path{arXiv:hep-th/9401072}}, \href
  {http://dx.doi.org/10.1016/0370-2693(94)91007-3}
  {\path{doi:10.1016/0370-2693(94)91007-3}}.

\bibitem{Holzhey:1994we}
Christoph Holzhey, Finn Larsen, and Frank Wilczek.
\newblock {Geometric and renormalized entropy in conformal field theory}.
\newblock {\em Nucl. Phys. B}, 424:443--467, 1994.
\newblock \href {http://arxiv.org/abs/hep-th/9403108}
  {\path{arXiv:hep-th/9403108}}, \href
  {http://dx.doi.org/10.1016/0550-3213(94)90402-2}
  {\path{doi:10.1016/0550-3213(94)90402-2}}.

\bibitem{Takayanagi:2017knl}
Tadashi Takayanagi and Koji Umemoto.
\newblock {Entanglement of purification through holographic duality}.
\newblock {\em Nature Phys.}, 14(6):573--577, 2018.
\newblock \href {http://arxiv.org/abs/1708.09393} {\path{arXiv:1708.09393}},
  \href {http://dx.doi.org/10.1038/s41567-018-0075-2}
  {\path{doi:10.1038/s41567-018-0075-2}}.

\bibitem{Synge:1960ueh}
J.~L. Synge.
\newblock {\em {Relativity: The General theory}}.
\newblock North-Holland., Amsterdam, The Netherlands, 1960.
\newblock \href {http://dx.doi.org/10.1126/science.132.3444.1933.c}
  {\path{doi:10.1126/science.132.3444.1933.c}}.

\end{thebibliography}

\end{document}